\documentclass[paper]{agujournal2019}

\usepackage{url}
\usepackage{lineno}
\usepackage{color}
\usepackage{amssymb,amsmath}
\usepackage[normalem]{ulem}
\usepackage{cancel}

\draftfalse

\journalname{JGR: Space Physics}

\begin{document}

\justifying


\title{Beam-driven Electron Cyclotron Harmonic and Electron Acoustic Waves as Seen in Particle-In-Cell Simulations}

\authors{Xu Zhang\affil{1}, Xin An\affil{1}, Vassilis Angelopoulos \affil{1}, Anton Artemyev \affil{1}, Xiao-Jia Zhang \affil{2}, Ying-Dong Jia \affil{1}}
\affiliation{1}{Department of Earth, Planetary, and Space Sciences, University of California, Los Angeles, USA}
\affiliation{2}{Department of Physics, University of Texas at Dallas, Richardson, TX, USA}

\correspondingauthor{Xu Zhang}{xuzhang@igpp.ucla.edu}

\begin{keypoints}
\item Electron cyclotron harmonic (ECH) and electron acoustic waves are excited by an electron beam in a 2-D particle-in-cell simulation.
\item Beam electrons are thermalized and cold electrons are accelerated through resonant interactions with ECH and electron acoustic waves.
\item When $\omega_{pe}/\omega_{ce}$ increases, ECH wave intensity increases while electron acoustic wave intensity decreases.
\end{keypoints}

\begin{abstract}
Recent study has demonstrated that electron cyclotron harmonic (ECH) waves can be excited by a low energy electron beam. Such waves propagate at moderately oblique wave normal angles ($\sim70^{\circ}$). The potential effects of beam-driven ECH waves on electron dynamics in Earth's plasma sheet is not known. Using two-dimensional Darwin particle-in-cell simulations with initial electron distributions that represent typical plasma conditions in the plasma sheet, we explore the excitation and saturation of such beam-driven ECH waves. Both ECH and electron acoustic waves are excited in the simulation and propagate at oblique wave normal angles. Compared with the electron acoustic waves, ECH waves grow much faster and have more intense saturation amplitudes. Cold, stationary electrons are first accelerated by ECH waves through cyclotron resonance and then accelerated in the parallel direction by both the ECH and electron acoustic waves through Landau resonance. Beam electrons, on the other hand, are decelerated in the parallel direction and scattered to larger pitch angles. The relaxation of the electron beam and the continuous heating of the cold electrons contribute to ECH wave saturation and suppress the excitation of electron acoustic waves. When the ratio of plasma to electron cyclotron frequency $\omega_{pe}/\omega_{ce}$ increases, the ECH wave amplitude increases while the electron acoustic wave amplitude decreases. Our work reveals the importance of ECH and electron acoustic waves in reshaping sub-thermal electron distributions and improves our understanding on the potential effects of wave-particle interactions in trapping ionospheric electron outflows and forming anisotropic (field-aligned) electron distributions in the plasma sheet.
\end{abstract}

\section{Introduction}\label{sec:1}
Electron cyclotron harmonic waves, known as $(n+1/2)f_{ce}$ waves, are electrostatic waves with dominant wave power in the frequency range between $f_{ce}$ and $2f_{ce}$ ($f_{ce}$ refers to electron cyclotron frequency) \cite{Kennel70,Fredricks71,Belmont83,Roeder&Koons89}. They are frequently observed in Earth's magnetosphere, confined near magnetic equator \cite{Meredith09:aurora}, extending over a large radial distance from the outer radiation belts \cite{Ni11:statistics, Ni17:ECH} to the plasma sheet \cite{Liang11:ECH,Zhang14:ECH,Zhang14:ECH&DF} and located in the night-side and dawn-side magnetosphere \cite{Thorne10:Nature,Ni11:statistics}. Electron cyclotron harmonic waves can resonate with electrons in the energy range between a few hundred eV to a few keV through cyclotron resonance when $\omega-k_{\parallel}v_{\parallel} = n|\omega_{ce}|$ is satisfied ($\omega$ is the wave frequency, $k_{\parallel}$ is the wave number in the direction parallel to the background magnetic field, $v_{\parallel}$ is the parallel velocity of resonant particle, $n$ is the resonance harmonic number and $\omega_{ce}=2\pi f_{ce}$) \cite{Lyons74:ECH:diffusion,Fontaine&Blanc83}. ECH waves, therefore, can scatter plasma sheet electrons into the loss cone and are the dominant driver of electron diffuse aurora precipitation in the outer magnetosphere beyond $8R_{E}$\cite{Horne&Thorne00,Horne03,Ni12:ECH,Zhang15:ECH}. ECH waves had long been thought to be driven unstable mainly by loss-cone distributions of plasma sheet electrons since early 1970s \cite{Young73,Karpman75:ECH,Maha&Kennel78,Maha79}. In a plasma environment consisting of two different electron populations: one cold component and one hot component, loss-cone driven ECH waves are unstable at wave normal angles $\theta(\vec{B},\vec{k}) = 88^{\circ} \sim 89^{\circ}$ and wave lengths comparable to the cold electron gyroradius $k_{\perp}\rho_{e,cold} \sim 1$ ($k_{\perp}$ is the perpendicular wavenumber and $\rho_{e,cold}$ is the cold electron gyroradius) \cite{Maha&Kennel78,Maha79}. Resonating with electrons at the velocity $v_{res} = (\omega-\omega_{ce})/k_{\parallel}$ comparable to the hot electron thermal velocity, ECH waves can be driven unstable by the loss cone distributions of hot electrons with $\partial{f}/\partial{v_{\perp}} > 0$ near the loss cone edge \cite{Horne89,Horne03,Ni12:ECH}. Excitation and saturation of loss-cone-driven ECH waves have been extensively investigated theoretically \cite{Liu18:ech,Liu20:ech}, observationally \cite{Zhang13:ECH} and with self-consistent particle-in-cell simulations \cite{Wu20:Tao:ECH}. Recent studies by \citeA{Zhang13:ECH} and \citeA{Wu20:Tao:ECH} demonstrate that loss-cone-driven ECH waves saturate through the filling of the loss cone. Waves can thus reach a marginally stable state between the wave growth by loss cone instabilities and filling of loss cone by wave scattering.

Apart from loss-cone-driven ECH waves at wave normal angles close to $90^{\circ}$, recent satellite observations demonstrate that there exists another population of ECH waves propagating at moderately oblique wave normal angles with respect to the background magnetic field ($\theta(\vec{B},\vec{k}) \sim 70^{\circ}$) in the plasma sheet \cite{Zhang21:jgr:ECH}. Enhancement of the parallel electron flux in the energy range of tens of eV to a few hundred eV coincides with observations of moderately oblique ECH waves, suggesting that these waves are likely to be driven unstable by low energy electron beams. The presence of low energy electron beams can significantly reduce the damping of ECH waves by Landau resonance with hot electrons, while free energy from cyclotron resonance with the electron beam can contribute to the generation of ECH waves at moderately oblique wave normal angles. Using a multi-component electron distribution function consisting of a cold (stationary) electron, beam electron and hot electron component, recent work has demonstrated that ECH waves can be driven unstable by the electron beam under a variety of plasma conditions. 
Both satellite observations \cite{Zhang21:jgr:ECH,Zhang22:ECH} and linear instability analysis \cite{Zhang21:pop:ECH} revealed the excitation of ECH waves by electron beam. However, the feedback of such beam-driven ECH waves on the electron plasma remains an open question. During the process of excitation and saturation of beam-driven ECH waves, the energy transfer between ECH waves and the different species of electrons needs to be addressed. This is the motivation for the present paper. Using a self-consistent particle-in-cell (PIC) simulation, we intend to confirm the generation of beam-driven ECH waves and investigate the feedback of those waves on the stationary cold, warm beam, and hot electron populations.

Various types of plasma waves can be generated by an electron beam including ECH waves mentioned above, whistler-mode waves \cite{Agapitov14:jgr:acceleration,Agapitov15:grl:acceleration,Mourenas15,Li16:statistics,Zhang22:natcom}, Langmuir waves \cite{Karpman75,Li17:langmuir,An19}, electron acoustic waves \cite{Gary&Tokar85,Gary87,Berthomier00,Vasko17:grl,Ma24}, etc. Wave-particle interaction in an electron-beam plasma has been extensively investigated in laboratory experiments \cite{VanCompernolle15,An16}, simulations \cite{Omura96,Gary00:beams,Lu05,An17:beam} and satellite observations \cite{Carlson98,Viberg13,Vasko17:grl}. Apart from ECH waves that are weakly damped in a plasma environment consisting of both cold (stationary) electron and hot electron as mentioned previously, electron acoustic waves, with their phase velocities higher than the thermal velocities of cold (stationary) electrons and yet lower than the thermal velocities of hot electrons, are also weakly damped in such a plasma environment. Consequently, these waves can be driven unstable by electron beams \cite{Watanabe&Taniuti77, Gary&Tokar85, Gary87, Mace91, Mace&Hellberg93}. During the excitation process of electron acoustic waves, nonlinear structures form in electron phase space, displaying characteristics of unipolar or bipolar electric fields \cite{Singh&Lakhina04, Lakhina11, Valentini06, Fu16:double_layers, Vasko17:grl, An19}. Such electric field signals, often termed as electrostatic solitary waves or time domain structures \cite{Malaspina14,Malaspina15,Mozer15}, have frequently been observed in the ionosphere 
\cite{Mozer77,Ergun98, Ergun01}, in the magnetotail \cite{Matsumoto94}, near magnetic reconnection sites \cite{Khotyaintsev10} and in other regions in space \cite{Cattell03,Malaspina13}. Electrostatic solitary waves can accelerate electrons at energies of tens of eV to energies of a few keV \cite{Artemyev14:grl:thermal, Mozer14}. They can also scatter electrons at energies below a few keV into the ionosphere and contribute to diffuse aurora precipitation \cite{Vasko17:diffusion,Shen20:tds}. In an electron-beam plasma system, various plasma waves can be excited and different wave modes compete with each other. Fastest growing plasma waves will grow first, extract free energy from the electron beam and therefore suppress the growing of other types of plasma waves \cite{Omura&Matsumoto87, An17:beam}. It is, therefore, worth addressing the relative importance between beam-driven ECH waves, beam-driven electron acoustic waves and other types of waves in a beam-plasma system. Our work here aims to improve our current understanding of the excitation of different plasma waves by an electron beam and the effects of different beam-driven waves on modifying the ambient electron distribution functions.  

In this work, we perform 2-D Darwin particle-in-cell simulations to investigate the interaction between beam-driven ECH waves and electron distributions and explore the competition between beam-driven ECH waves with other plasma waves in a beam-plasma environment. Section \ref{sec:2} describes the initial simulation setup. In Section \ref{sec:3}, we demonstrate the excitation of beam-driven ECH waves and electron acoustic waves and explore their properties. The evolution of different electron populations as a result of wave-particle interactions is analyzed in Section \ref{sec:4}. Section \ref{sec:5} discusses the competition between beam-driven ECH waves and electron acoustic waves under different plasma environments. Section \ref{sec:6} is our summary and discussion.

\section{Simulation Setup}\label{sec:2}
The Darwin particle-in-cell code used in this work is an electromagnetic PIC code with periodic boundary conditions, 2D in configuration space and 3D in velocity space, developed as part of the UCLA particle-in-cell Framework \cite{Decyk07}. In a conventional electromagnetic PIC code, electric and magnetic fields are solved using Maxwell's equations. The time step, $\Delta t$, the grid length, $\Delta$, and the number of spatial dimensions, $N_{d}$, must satisfy the Courant condition $c\Delta t \le \Delta/N_{d}$ so that a light wave can be resolved in such a simulation. In a Darwin PIC code, the transverse electric field $\bold E_{T}$ is neglected in Ampere's law, which is expressed as 

\begin{equation}
\nabla\times\bold{B}=\frac{4\pi}{c}\bold{j}+\frac{1}{c}\frac{\partial\bold{E}_{L}}{\partial t}
\end{equation}
\label{eq01}
where $\bold{E}_{L}$ is the longitudinal electric field, and $\bold E_{T}$ and $\bold E_{L}$ are defined as $\bold{k}\cdot\bold{E}_{T} = 0$ and $\bold{k}\times\bold{E}_{L} = 0$. Because light waves are eliminated in the Darwin model, the time step does not need to be very small in order to satisfy the Courant condition and it is chosen to be $\Delta t = 0.02\omega^{-1}_{pe}$ in our simulation. We use a simulation box with $1024\times1024$ grids in x and y directions. The grid size is $\Delta = 0.002c/\omega_{pe}$, where $\omega_{pe}$ is the electron plasma frequency and $c$ is the speed of light. Time is normalized to $\omega^{-1}_{pe}$ and velocity is normalized to $v_{0} = 0.002c$. The background magnetic field is directed along x with $\omega_{pe}/\omega_{ce} = 5$ in our simulation.

The initial electron distribution functions in our simulation are drifting Maxwellians:

\begin{equation}
f(v_{\perp},v_{\parallel}) = \frac{n}{(\sqrt{2\pi})^{3}{v}^2_{th\perp}v_{th\parallel}}e^{-\frac{(v_{\parallel}-v_{d})^2}{2{v}^2_{th\parallel}}}e^{-\frac{v^2_{\perp}}{2v^2_{th\perp}}}
\end{equation}
\label{eq02}
where $n$ is the number density, $v_{th\perp}$ and $v_{th\parallel}$ are thermal velocities in directions perpendicular and parallel to the background magnetic field and $v_{d}$ is the drift velocity. They consist of three components: (1) cold electrons with thermal velocity $v_{th,cold} = v_{0}$; (2) hot electrons with parallel thermal velocity $v_{th,hot,\parallel} = 31.62v_{0}$ and temperature anisotropy $T_{\perp}/T_{\parallel} = 0.85$; and (3) beam electrons with parallel thermal velocity $v_{th,beam,\parallel} = 10v_{0}$, temperature anisotropy $T_{\perp}/T_{\parallel} = 0.85$ and parallel drift velocity $v_{drift} = 30v_{0}$. The density ratios between these three components are $n_{cold}:n_{hot}:n_{beam} = 1:10:1$. There parameters represent the typical plasma parameters near the ECH wave generation region in Earth's magnetotail \cite{Zhang21:jgr:ECH, Zhang22:ECH}. In our simulations, there are 64 marker particles for hot electrons, 64 marker particles for cold electrons and 64 marker particles for beam electrons in each cell. Each electron component is assigned a weight to represent their respective densities. We decreased the charge $q$ of each particle and increased the number of particles $N$ in each cell for cold and beam electron species.This adjustment maintains the quantity $qN$ constant, ensuring that the particle dynamics remain unaltered. Ions are immobile in our simulations.

\section{Wave properties}\label{sec:3}
\begin{figure} 
\centering
\includegraphics[width=1\textwidth]{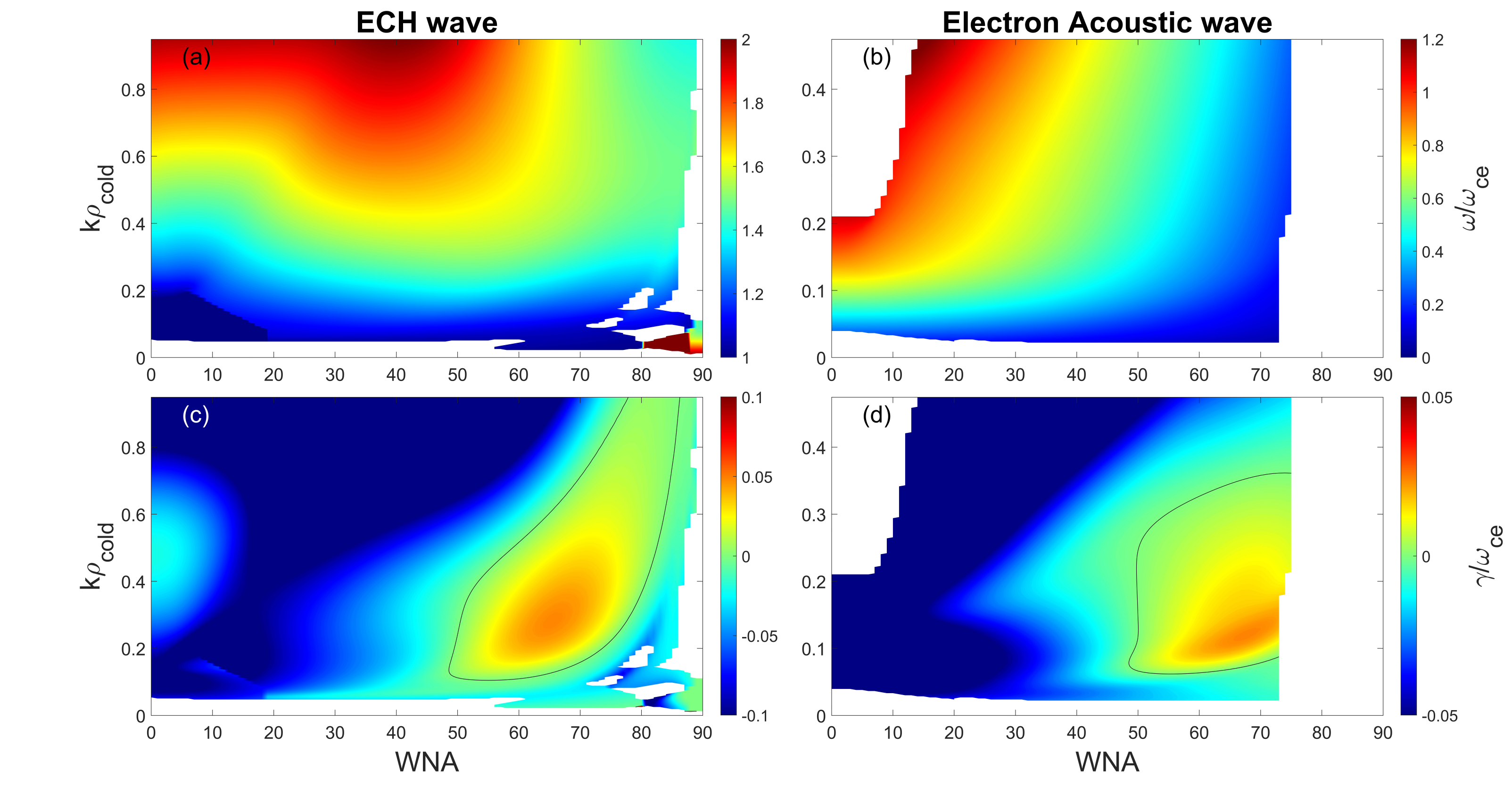}
\caption{Dispersion relation of ECH waves and electron acoustic waves. (a): wave frequency of ECH waves as a function of normalized wave vector and wave normal angle, where frequency $\omega$ is normalized to the electron cyclotron frequency $\omega_{ce}$ and the wave vector is normalized to the cold electron gyroradius $\rho_{e,cold}$; (b): wave frequency of electron acoustic waves; (c): growth rate of ECH waves, growth rate $\gamma$ is normalized to $\omega_{ce}$; (d): growth rate of electron acoustic waves.}
\label{fig1}
\end{figure}

Before analyzing simulation results on wave-particle interactions in the beam-plasma system, we first examine the linear dispersion relation of plasma waves driven by an electron beam using WHAMP (Waves in Homogeneous, Anisotropic Multi-component Plasmas) \cite{Roennmark82}. Considering a background magnetic field of 50$nT$, electron parameters discussed in Section \ref{sec:2}, and typical electron density and temperature parameters of the plasma sheet, the initial electron distributions in our simulation expressed in real physical units are shown in Table 1. Using electron distribution functions in Table 1 as input to WHAMP, dispersion relations for ECH waves in the first harmonic frequency band and electron acoustic waves are derived and plotted in Figure \ref{fig1}. We see that both ECH and electron acoustic waves are driven unstable by the field-aligned electron beam and propagate at oblique angles with respect to the ambient magnetic field. Although the wave properties of the unstable wave modes shown in Figures \ref{fig1}(b) and \ref{fig1}(d) and the electrostatic waves in the whistler frequency range as seen in our simulations resemble the wave properties of oblique whistler-mode waves \cite{Artemyev15:natcom,Mourenas12:JGR,Artemyev22:jgr:Landau&ELFIN}, these oblique propagating electrostatic waves are in fact electron acoustic waves (see supporting information for detailed discussions). White patches at small wave normal angles in Figures \ref{fig1}(b) and \ref{fig1}(d) are attributed to the strong damping and white patches at large wave normal angles in Figures \ref{fig1}(b) and \ref{fig1}(d) arise from the fact that electron acoustic waves cannot propagate at such large wave normal angles\cite{Mace&Hellberg93}. Beam electrons can resonate with ECH and electron acoustic waves and provide free energy for wave generation. The resonance condition is:

\begin{equation}
\omega-k_{\parallel}v_{\parallel}=n|\omega_{ce}|
\end{equation}
\label{eq03}
where $n = 0,\pm1,\pm2...$ is the resonance harmonic number. The most unstable ECH wave is at the wave frequency $f/f_{ce} = 1.45$ and wave normal angle $\theta(\vec{B},\vec{k})=65^{\circ}$. Moderately oblique ECH waves are driven unstable through Landau resonance with the electron beam when $n=0$ and through cyclotron resonance when $n=-1$ and $n=-2\sim-5$ \cite{Zhang21:pop:ECH}. The most unstable electron acoustic wave is at wave frequency $f/f_{ce} = 0.23$ and wave normal angle $\theta(\vec{B},\vec{k})=66^{\circ}$ and is driven unstable through both Landau resonance and cyclotron resonance with the electron beam. 

\begin{figure} 
\centering
\includegraphics[width=1\textwidth]{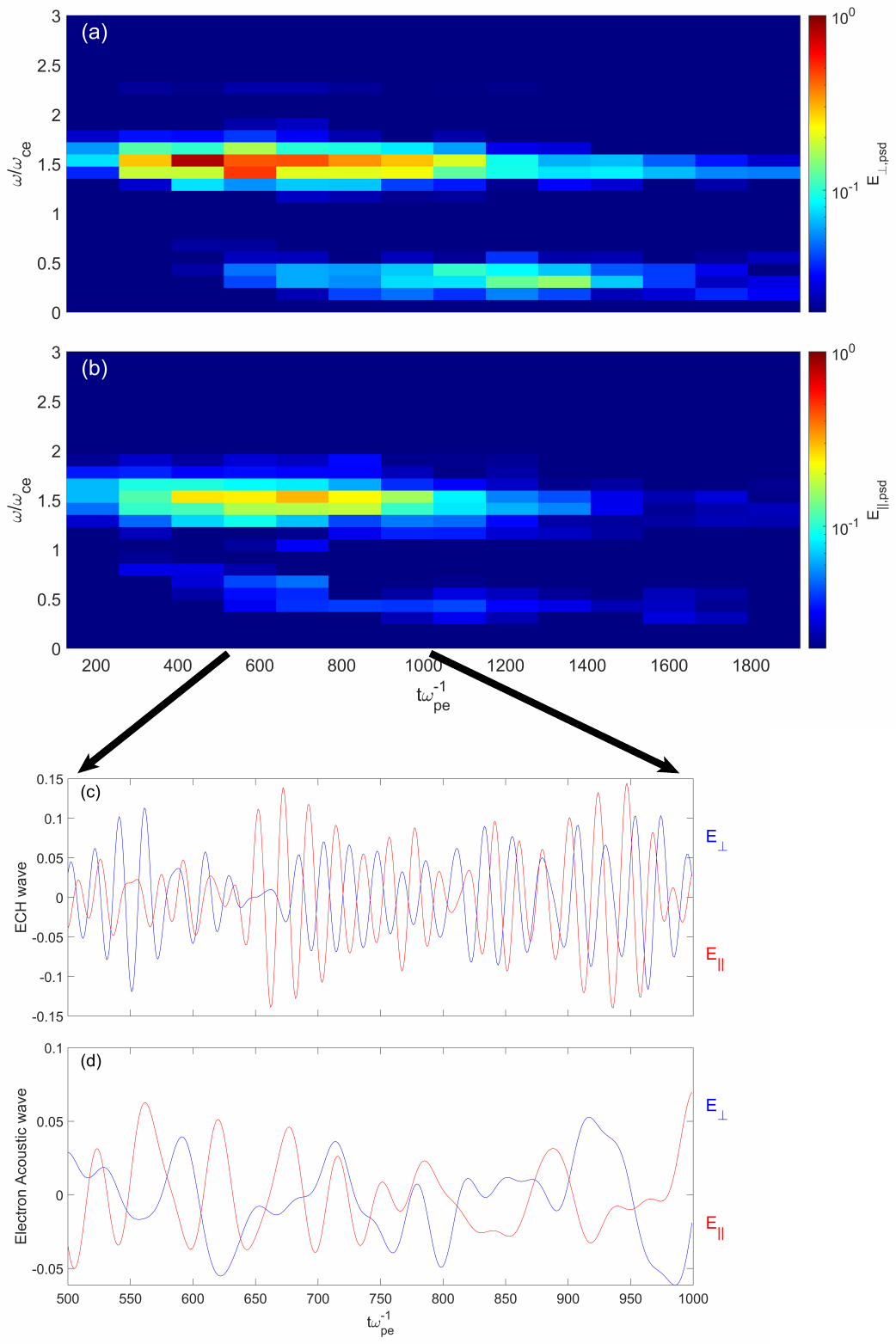}
\caption{Dynamic power spectrum of perpendicular and parallel component of longitudinal electric field $\vec{E}_{L}$. (a): power spectral density of $E_{\perp}$ as a function of frequency and time. Frequency is normalized to $\omega_{ce}$ and time is normalized to $\omega^{-1}_{pe}$; (b): power spectral density of $E_{\parallel}$; (c): $E_{\perp}$ and $E_{\parallel}$ at $x = 512\Delta$ and $y = 512\Delta$ after bandpass filtering in the frequency range of $[f_{ce}, 2f{ce}]$; (d): $E_{\perp}$ and $E_{\parallel}$ at $x = 512\Delta$ and $y = 512\Delta$ after bandpass filtering in the frequency range of $[0.05f_{ce}, f{ce}]$.}
\label{fig2}
\end{figure}

Figure \ref{fig2} shows the dynamic power spectrum of longitudinal electric field $\vec{E}_{L}$ in our simulation. $\vec{E}_{L}$ is diagnosed every $t=1\omega^{-1}_{pe}$ in each cell. Its perpendicular and parallel components are Fourier-transformed to frequency domain with a time window of $256\omega^{-1}_{pe}$. The power spectral densities of $E_{\perp}$ and $E_{\parallel}$ are averaged in space over 1024 cells along x direction at $y = 512\Delta$ and over 1024 cells along y direction at $x = 512\Delta$. The averaged power spectra are shown in Figures \ref{fig2}(a) and \ref{fig2}(b). After bandpass filtering with the frequency range in $[f_{ce}, 2f{ce}]$ and in $[0.05f_{ce}, f_{ce}]$, $E_{\perp}$ and $E_{\parallel}$ at the location $x = 512\Delta$ and $y = 512\Delta$ are shown in Figures \ref{fig2}(c) and \ref{fig2}(d) respectively. It is evident that both ECH and electron acoustic waves are excited in our simulation. ECH waves grow first and reach peak power at the frequency $f/f_{ce} = 1.47$ and at around $t=400\sim500\omega^{-1}_{pe}$, while electron acoustic waves grow much slower with peak power at a frequency $f/f_{ce} = 0.245$ and at around $t=1200\sim1400\omega^{-1}_{pe}$. The ECH waves carry a strong parallel electric field comparable to the perpendicular one, suggesting that the wave normal angle is much smaller than $90^{\circ}$. The electron acoustic wave is also propagating at oblique wave normal angles as indicated by both perpendicular and parallel components of the wave electric field.

\begin{figure} 
\centering
\includegraphics[width=1\textwidth]{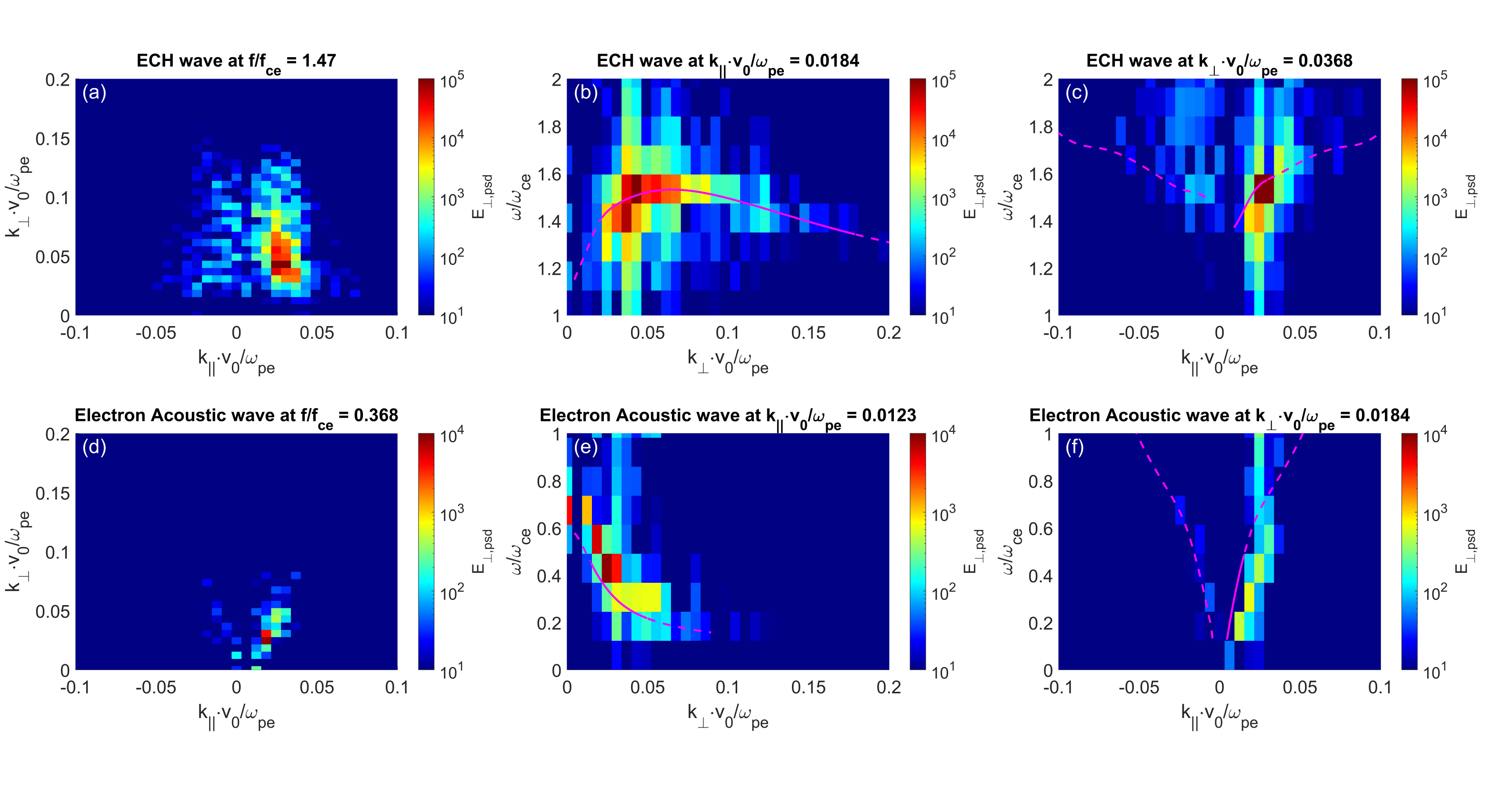}
\caption{(a): Power spectral density of $E_{\perp}$ as a function of normalized wave vector in perpendicular direction and in parallel direction at wave frequency $\omega/\omega_{ce} = 1.473$. $k_{\perp}$ and $k_{\parallel}$ are normalized to the quantity $v_{0}/\omega_{pe}$; (b): Power spectral density of $E_{\perp}$ as a function of $k_{\perp}v_{0}/\omega_{pe}$ and $\omega/\omega_{ce}$ at $k_{\parallel}v_{0}/\omega_{pe} = 0.0184$. Magenta line indicates linear dispersion relation of ECH wave solved using WHAMP. Solid magenta line represents wave growth and dashed magenta line represents wave damping from linear instability analysis; (c): Power spectral density of $E_{\perp}$ as a function of $k_{\parallel}v_{0}/\omega_{pe}$ and $\omega/\omega_{ce}$ at $k_{\perp}v_{0}/\omega_{pe} = 0.0368$. Magenta line indicates linear dispersion relation of ECH wave; (d): Power spectral density of $E_{\perp}$ as a function of $k_{\perp}v_{0}/\omega_{pe}$ and $k_{\parallel}v_{0}/\omega_{pe}$ at $\omega/\omega_{ce} = 0.368$; (e): Power spectral density of $E_{\perp}$ as a function of $k_{\perp}v_{0}/\omega_{pe}$ and $\omega/\omega_{ce}$ at $k_{\parallel}v_{0}/\omega_{pe} = 0.0123$. Magenta line indicates linear dispersion relation of electron acoustic wave; (f): Power spectral density of $E_{\perp}$ as a function of $k_{\parallel}v_{0}/\omega_{pe}$ and $\omega/\omega_{ce}$ at $k_{\parallel}v_{0}/\omega_{pe} = 0.0184$. }
\label{fig3}
\end{figure}

The perpendicular wave electric field $\delta E_{y}(x,y,t)$ in the time window between $384\omega^{-1}_{pe}$ and $640\omega^{-1}_{pe}$ is Fourier-transformed in $\omega-k_{x}-k_{y}$ space (background magnetic field in x direction). The electric field power in the ECH frequency range maximizes at $f/f_{ce} = 1.473$, $k_{x}v_{0}/\omega_{pe} = 0.0184$ and $k_{y}v_{0}/\omega_{pe} = 0.0368$ as shown in Figure \ref{fig3}(a), corresponding to a wave normal angle of $63^{\circ}$. The electric field power spectral density at $k_{x}v_{0}/\omega_{pe} = 0.0184$ and at $k_{y}v_{0}/\omega_{pe} = 0.0368$ are plotted in Figures \ref{fig3}(b) and \ref{fig3}(c) respectively and their observed dispersion agrees with the linear instability analysis. The electric field power in the frequency range below $f_{ce}$ maximizes at $f/f_{ce} = 0.368$, $k_{x}v_{0}/\omega_{pe} = 0.0123$ and $k_{y}v_{0}/\omega_{pe} = 0.0184$, suggesting that the electron acoustic wave propagates at $\theta(\vec{B},\vec{k}) = 56^{\circ}$. The electric field power spectral density at $k_{x}v_{0}/\omega_{pe} = 0.0123$ and at $k_{y}v_{0}/\omega_{pe} = 0.0184$ are shown in Figures \ref{fig3}(e) and \ref{fig3}(f). 

\section{Electron dynamics}\label{sec:4}
\begin{figure} 
\centering
\includegraphics[width=1\textwidth]{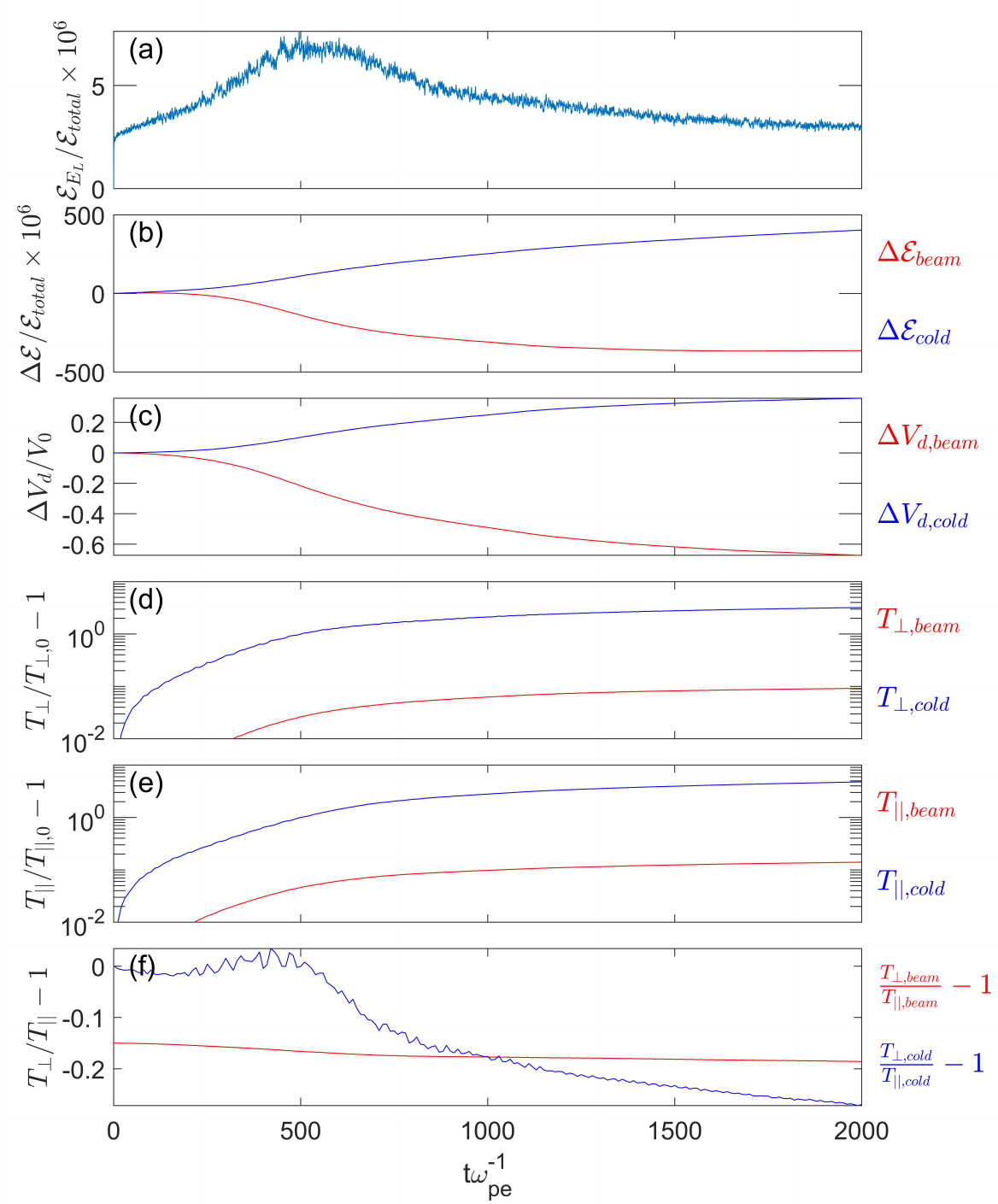}
\caption{(a): $\mathcal{E}_{E_{L}}/\mathcal{E}_{total}\times10^6$. $\mathcal{E}_{E_{L}}$ is energy of longitudinal electric field, $\mathcal{E}_{total}$ is the sum of electric and magnetic fields and particle kinetic energies in the simulation; (b):$\Delta\mathcal{E}$ is the difference between the electron kinetic energy at time $t$ and at $t=0$. Red color represents the beam and blue color represents the cold electrons; (c): Change in drift velocity of electron beam and cold electron components; (d): Change in the perpendicular temperature of beam and cold electron components; (e): Change in the parallel temperature of beam and cold electron components; (f): Temperature anisotropy of beam and cold electron components.}
\label{fig4}
\end{figure}

Section \ref{sec:3} demonstrates that both ECH and electron acoustic waves are excited by an electron beam and propagate at oblique wave normal angles with respect to the ambient magnetic field. In this section, we will investigate the effects of these two waves on electron dynamics. The time evolution of the electron kinetic energy and electron temperature are illustrated in Figure 4. The wave power of moderately oblique ECH wave increases at the start of simulation, reaches its maximum at around $t=500\omega^{-1}_{pe}$ and gradually decreases afterwards (see Figure \ref{fig2}(a) and Figure \ref{fig4}(a)). The ECH wave gains free energy from the beam electrons; in the meantime, the ECH wave transfers some of its energy to the kinetic energy of cold electrons. The oblique electron acoustic wave, with slower wave growth and weaker intensity compared with ECH wave, also gains energy from beam electrons during wave excitation.The total energy, which is the summation of the energy of electric fields, magnetic fields and electrons, is conserved in the simulation. Resonating with the ECH and electron acoustic wave through Landau resonance with $v_{res} = \omega/k_{\parallel}$, the electron beam is thermalized in the parallel direction and its drift velocity decreases. The ECH and electron acoustic waves can also resonate with beam electrons through cyclotron resonance and accelerate beam electrons in the perpendicular direction. The cold electrons are continuously accelerated in both the parallel and perpendicular directions. In the beginning of our simulation, before $t\sim500\omega^{-1}_{pe}$, the heating of cold electrons in the perpendicular direction is more effective than the heating in the parallel direction, resulting in a slightly perpendicular anisotropy. Later in our simulation, the cold electrons are accelerated preferentially in the parallel direction. 

\begin{figure} 
\centering
\includegraphics[width=1\textwidth]{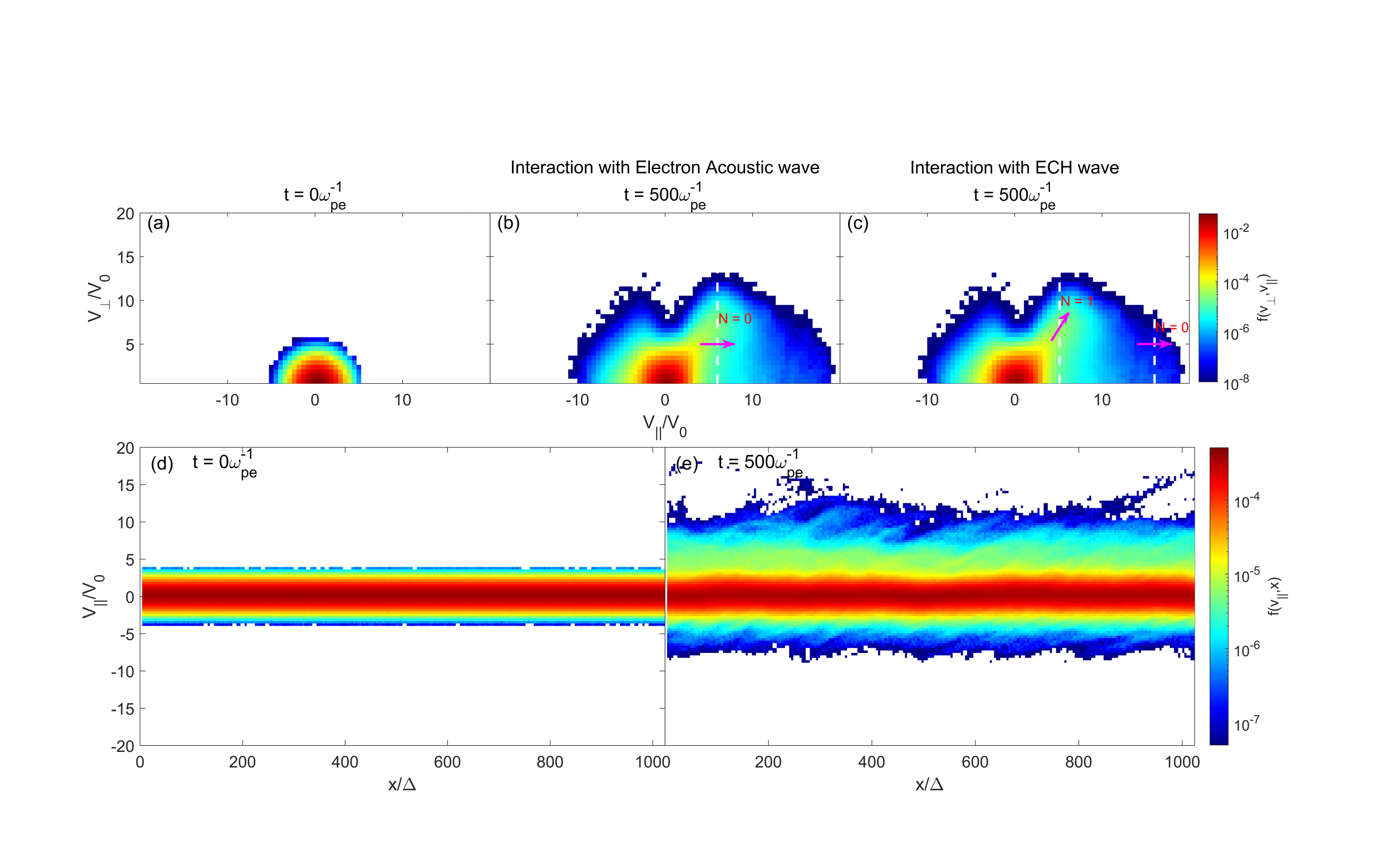}
\caption{(a): Cold electron phase space density as a function of $v_{\perp}/v_{0}$ and $v_{\parallel}/v_{0}$ at $t=0\omega^{-1}_{pe}$; (b): Cold electron phase space density at $t=500\omega^{-1}_{pe}$. White dashed line indicates Landau resonance velocity with oblique electron acoustic wave at $v_{res} = 5.98v_{0}$. Magenta line indicates the diffusion curve and arrow represents the direction in which particles diffuse; (c): Cold electron phase space density at $t=500\omega^{-1}_{pe}$. White dashed lines indicate Landau resonance velocity ($v_{res} = 16v_{0}$) and cyclotron resonance velocity ($v_{res} = 5.14v_{0}$) with ECH wave. Magenta lines represent diffusion curves for Landau and cyclotron resonance with ECH wave; (d): Cold electron phase space density as a function of $v_{x}$ and $x$ at $t=0\omega^{-1}_{pe}$; (e): Cold electron phase space density as a function of $v_{x}$ and $x$ at $t=500\omega^{-1}_{pe}$.}
\label{fig5}
\end{figure}

The cold electron distribution function at $t=500\omega^{-1}_{pe}$ in Figure \ref{fig5}(b) demonstrates that some of the cold electrons are accelerated in both parallel and perpendicular directions. The beam-driven electron acoustic wave resonates with cold electrons through Landau resonance with $v_{res} = 5.98v_{0}$ as illustrated in Figure \ref{fig5}(b). The parallel electric field of the electron acoustic wave, however, is not very strong before $t=500\omega^{-1}_{pe}$ (see Figure \ref{fig2}(b)). The oblique electron acoustic wave, therefore, is not very effective in acceleration of cold electrons at the start of simulation. The beam-driven ECH wave, with strong wave electric field in both perpendicular and parallel directions, can resonate with the cold electrons through cyclotron resonance with $v_{res} = (\omega-|\omega_{ce}|)/k_{\parallel} = 5.14v_{0}$ and Landau resonance with $v_{res} = 16v_{0}$. To analyze the resonant interactions between ECH waves and cold electrons, diffusion curves are illustrated in Figure \ref{fig5}(c) and defined as

\begin{equation}
C_{n} = nW-\mu\omega = constant
\end{equation}
\label{eq04}
where $W$ is the particle kinetic energy, $\mu=(\frac{1}{2}mv^2_{\perp})/|\omega_{ce}|$, $n=0$ indicates Landau resonance and $n=1$ indicates cyclotron resonance \cite{Shklyar09:review}. In the beginning of the simulation, the ECH wave resonates with the high energy tail of the cold electron distribution through cyclotron resonance and accelerates cold electrons in both perpendicular and parallel directions (see Figures \ref{fig4}(f) and Figure \ref{fig5}(c)). When cold electrons have already been heated through cyclotron resonance and have enough particles at velocities around $v = 10\sim15v_{0}$, the ECH wave can accelerate those cold electrons in the parallel direction through Landau resonance. Figure \ref{fig5}(e) shows the cold electron phase space density in $v_{x}-x$ space, with clear signatures of non-linear Landau trapping of the high energy tail of the cold electron distributions at around $v = 10\sim15v_{0}$. Considering a monochromatic wave with parallel wave electric field $\delta E_{\parallel}$, the half-width of the trapping island in $v_{\parallel}$ is 

\begin{equation}
\delta v_{\parallel} = \sqrt{\frac{2e\delta\phi}{m}} = \sqrt{\frac{2\delta E_{\parallel}}{k_{\parallel}m}}
\end{equation}
\label{eq05}
Taking into account the wave amplitude of the ECH wave at $(e\delta E_{\parallel})/(m\omega_{pe}v_{0}) = 0.15$ and wave number at $k_{\parallel}v_{0}/\omega_{pe} = 0.0184$, the beam-driven ECH wave resonates with cold electrons at $v_{res} = 16$ with a half-width at $\delta v_{\parallel} = 4v_{0}$. Therefore, the ECH wave can trap some of the particles at the tail of the cold electron distribution and accelerate them in the parallel direction through nonlinear Landau resonance. 

\begin{figure} 
\centering
\includegraphics[width=1\textwidth]{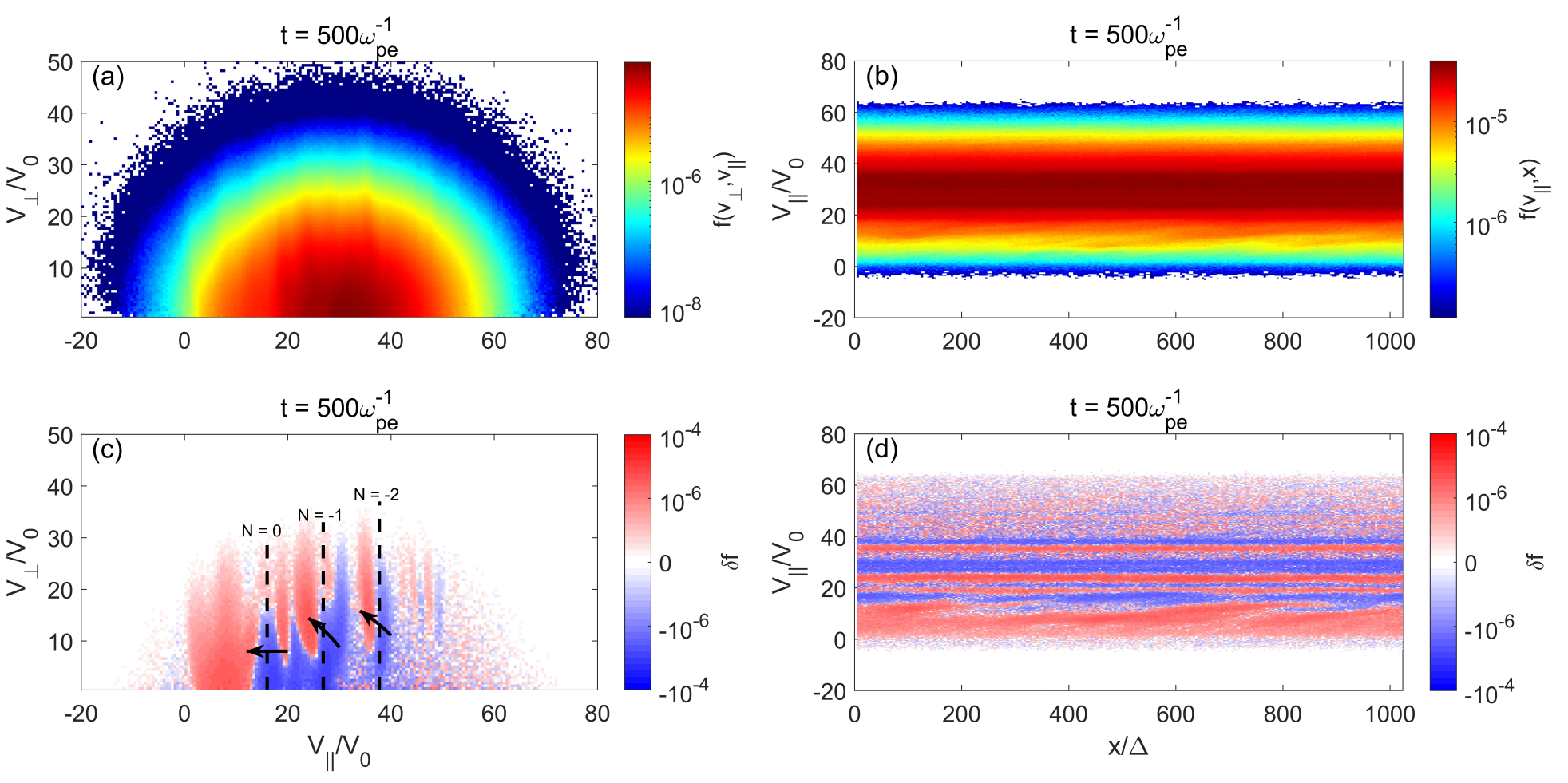}
\caption{(a): Beam electron phase space density as a function of $v_{\perp}/v_{0}$ and $v_{\parallel}/v_{0}$ at $t=500\omega^{-1}_{pe}$; (b): Beam electron phase space density as a function of $v_{x}$ and $x$ at $t=500\omega^{-1}_{pe}$; (c): Difference in beam electron phase space density in velocity space. $\delta f(v_{\perp},v_{\parallel}) = f(v_{\perp},v_{\parallel}, t = 500\omega^{-1}_{pe}) - f(v_{\perp},v_{\parallel}, t = 0\omega^{-1}_{pe})$. Red color indicates increase in electron phase space density and blue color indicates decrease in electron phase space density. Dashed black lines indicate resonance velocities with ECH wave at Landau resonance when $n=0$ and cyclotron resonances when $n = -1,-2$. Solid black lines indicate diffusion curves and black arrows represent directions in which particles diffuse. (d): $\delta f(v_{x},x) = f(v_{x},x, t = 500\omega^{-1}_{pe}) - f(v_{x},x, t = 0\omega^{-1}_{pe})$. }
\label{fig6}
\end{figure}

Figure \ref{fig6} shows the beam electron distribution function at $t=500\omega^{-1}_{pe}$ in $v_{\perp}-v_{\parallel}$ space and in $v_{\parallel}-x$ space. By conserving $\mu=(\frac{1}{2}mv^2_{\perp})/|\omega_{ce}|$, a beam electron can exchange energy with the ECH wave in the parallel direction through Landau resonance and the net energy transport direction is directed towards ECH wave growth. The ECH wave can also scatter the beam electron to larger pitch angles through cyclotron resonance when $n=-1$ at $v_{res} = 26.9v_{0}$ and $n=-2$ at $v_{res}=37.8v_{0}$. Cyclotron resonance with the ECH wave moves the beam electron along the diffusion curve and the direction of net particle transport is depicted by black arrows in Figure \ref{fig6}(c). Kinetic energy of resonating electrons is transferred to ECH waves while the electron perpendicular energy increases. Resulting from resonant wave-particle interactions, structures are formed in velocity space of the beam electron distribution function in Figures \ref{fig6}(a) and \ref{fig6}(c). Signatures of non-linear Landau trapping of beam electrons by the ECH waves at velocities around $v = 10\sim15v_{0}$ are demonstrated in Figures \ref{fig6}(b) and \ref{fig6}(d). The electron acoustic wave, with Landau resonance velocity at $v_{res} = 5.98v_{0}$ and cyclotron resonance velocity at $v_{res} = 22.2v_{0}$ when $n=-1$, can also thermalize beam electrons in the parallel direction and scatter beam electrons to larger pitch angles.

\section{Competition between beam-driven ECH wave and beam-driven electron acoustic wave}\label{sec:5}
\begin{figure} 
\centering
\includegraphics[width=1\textwidth]{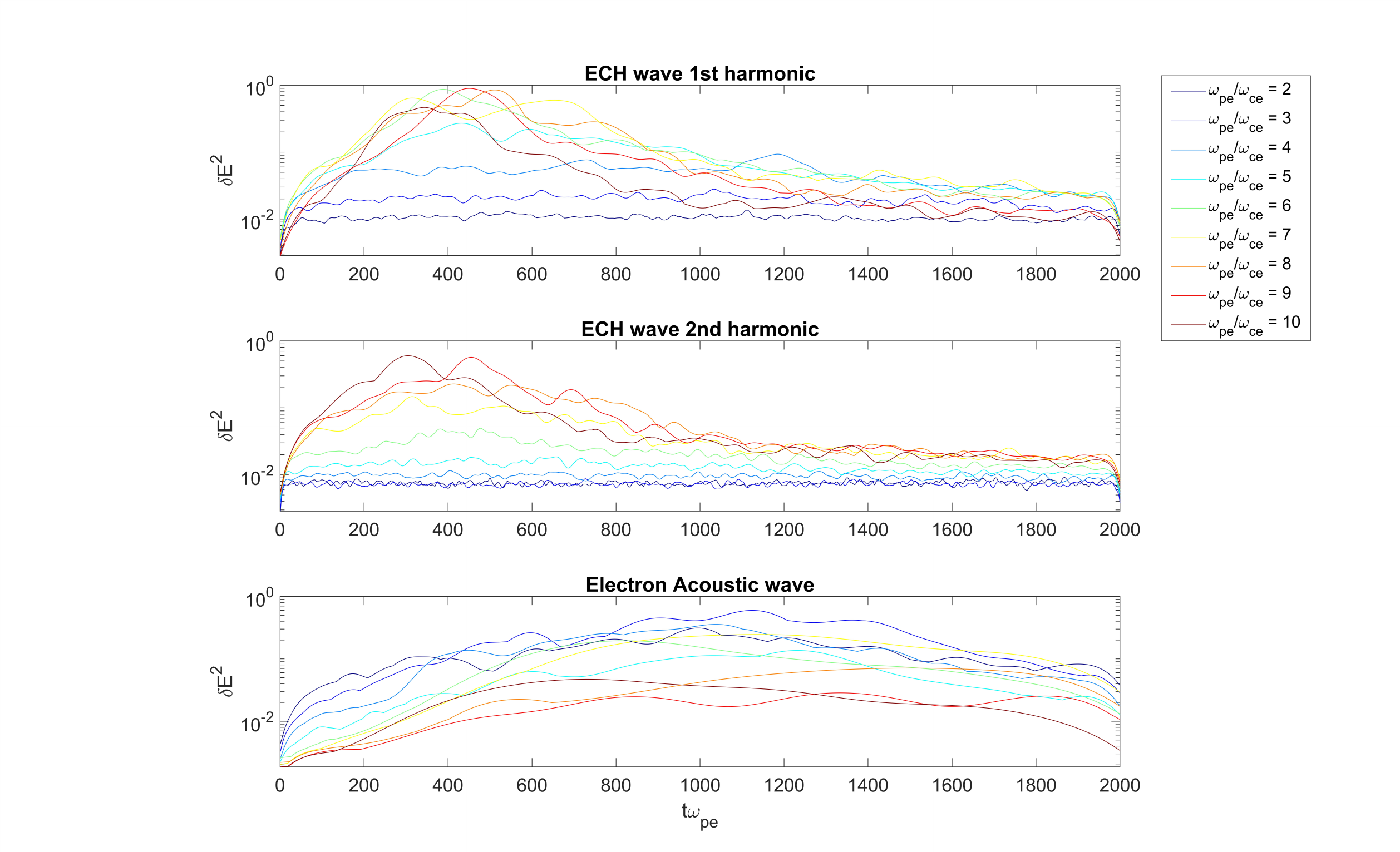}
\caption{(a): Electric field wave power in the frequency range $[f_{ce},2f_{ce}]$. Different colors represent different $\omega_{pe}/\omega_{ce}$ ratios; (b): Electric field wave power in the frequency range $[2f_{ce},3f_{ce}]$; (c): Electric field wave power in the frequency range $[0.05f_{ce},f_{ce}]$. }
\label{fig7}
\end{figure}

\begin{figure} 
\centering
\includegraphics[width=0.8\textwidth]{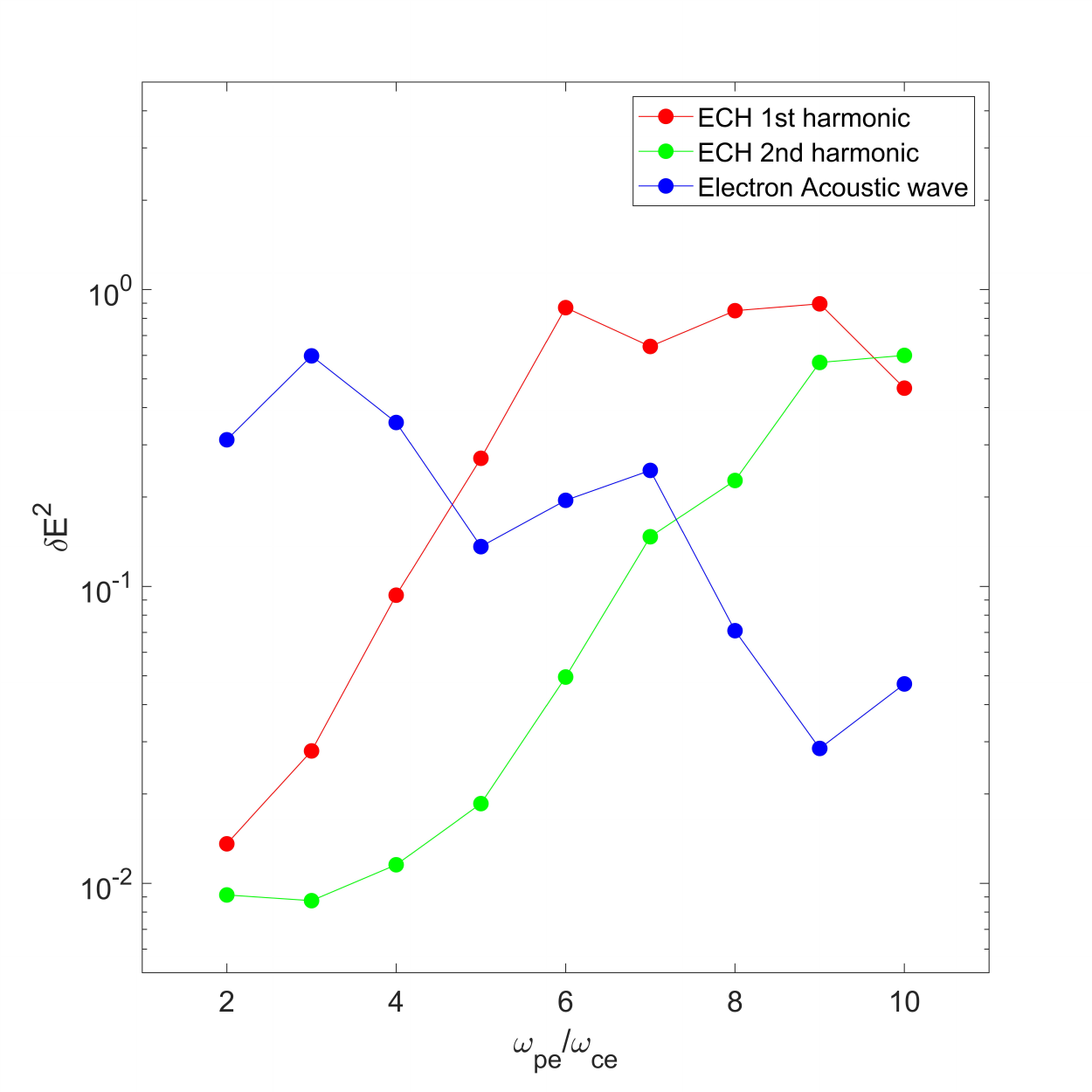}
\caption{Maximum electric field wave power for the first harmonic ECH wave, the second harmonic ECH wave and oblique electron acoustic wave as a function of $\omega_{pe}/\omega_{ce}$ ratio}
\label{fig8}
\end{figure}

The moderately oblique ECH waves gain energy from the electron beam and are excited first in our simulation. After reaching peak wave power at around $t=500\omega^{-1}_{pe}$, the ECH waves saturate and their intensity gradually decreases. Heating of the cold electrons through resonant wave-particle interactions and the subsequent relaxation of the electron beam contribute significantly to the saturation of beam-driven ECH waves. Beam-driven electron acoustic waves grow much slower and to a much weaker intensity compared with ECH waves. Because the free energy in the electron beam is depleted and the waves can be damped by cold electrons through Landau resonance, the growth of oblique electron acoustic waves is suppressed by the ECH waves. The time scales on the excitation process of the ECH and electron acoustic waves, which are determined by their linear growth rates, play an important role in the competition between beam-driven ECH and electron acoustic waves. The ratio between the electron plasma frequency and electron cyclotron frequency, $\omega_{pe}/\omega_{ce}$, is another important parameter that controls the linear growth rates of these two waves. To examine these relationships we transform the electrostatic wave field $\vec E_{L}$ into the frequency domain using wavelet analysis \cite{torrence&compo98} and average its dynamic power spectra over 1024 cells along x direction when $y = 512\Delta$ and over 1024 cells along y direction when $x = 512\Delta$. The time evolution of the peak electric field wave power in the frequency range of $[f_{ce},2f_{ce}]$, $[2f_{ce},3f_{ce}]$ and $[0.05f_{ce},f_{ce}]$ is shown in Figure \ref{fig7} for different $\omega_{pe}/\omega_{ce}$ ratios. Figure \ref{fig8} shows the maximum electric field wave power for ECH waves in the first harmonic frequency band, for ECH waves in the second harmonic frequency band and for oblique electron acoustic waves, as a function of $\omega_{pe}/\omega_{ce}$. When $\omega_{pe}/\omega_{ce}$ increases, the linear stability analysis predicts that growth rate of the first harmonic ECH wave increases (see supporting information for the dependence of linear growth rates on $\omega_{pe}/\omega_{ce}$ for different waves) and thus the wave intensity increases. The second harmonic ECH wave becomes unstable when $\omega_{pe}/\omega_{ce}>6$ and its intensity increases with increasing $\omega_{pe}/\omega_{ce}$ ratio. We notice that the intensity of the first harmonic ECH wave saturates when $\omega_{pe}/\omega_{ce}$ is larger than 6, suggesting that the growth of the second harmonic ECH wave restrains the intensity at its first harmonic frequency band. Results from this linear instability analysis suggest that linear growth rate of oblique electron acoustic waves should increase with increasing $\omega_{pe}/\omega_{ce}$ ratio. The intensity of the oblique electron acoustic wave, however, decreases as $\omega_{pe}/\omega_{ce}$ increases. The oblique electron acoustic wave, with a slower growth rate than the ECH wave, is suppressed due to the faster growth of the ECH wave. The intensity of the oblique electron acoustic wave, therefore, is anti-correlated with the intensity of beam-driven ECH wave and decreases with increasing $\omega_{pe}/\omega_{ce}$ ratio.

\section{Summary and Discussion}\label{sec:6}
In our work, we use 2D Darwin PIC simulations to explore the excitation and saturation of beam-driven ECH waves and beam-driven electron acoustic waves under plasma conditions in Earth's plasma sheet. Excited by Landau resonance and cyclotron resonance with beam electrons, the ECH wave grows faster and has stronger intensity while the electron acoustic wave grows more slowly and has weaker intensity. Both ECH and electron acoustic waves in our simulations are characterized by parallel wave electric field and propagate in oblique wave normal angles at around $65^{\circ}$. Free energy is transferred from the beam electrons to the ECH and electron acoustic waves and these two waves can efficiently accelerate cold electrons. Cold electrons are thus heated in the perpendicular and parallel directions through cyclotron resonance in the beginning of the simulation and then accelerated in the parallel direction through Landau resonance with both the ECH and the electron acoustic waves. The parallel temperature of the cold electrons increases by a factor of 6 and its perpendicular temperature increases by a factor of 4 at $t=2000\omega^{-1}_{pe}$. ECH and electron acoustic waves resonate with beam electrons through Landau resonance and gradually flatten the positive phase space density slope in the parallel direction. They can also scatter beam electrons to larger pitch angles through cyclotron resonance. Eventually, both ECH and electron acoustic waves saturate due to exhaustion of the free energy in beam electrons and the continuous heating of the cold electrons. We also investigate the competition of beam-driven ECH waves and beam-driven electron acoustic waves under different $\omega_{pe}/\omega_{ce}$ ratios. The wave power of the first harmonic ECH wave and the second harmonic ECH wave increase with increasing $\omega_{pe}/\omega_{ce}$ ratio. The wave power of the electron acoustic wave decreases with increasing $\omega_{pe}/\omega_{ce}$ ratio. Previous theoretical work in \citeA{Zhang21:jgr:ECH,Zhang21:pop:ECH} focused on linear instability analysis of excitation of beam-driven ECH waves. This study, however, not only investigates the excitation of ECH waves but also analyzes the effects of ECH waves on modifying the electron distribution function. ECH waves can resonate with cold electrons and beam electrons through nonlinear Landau trapping and accelerate or decelerate them in the parallel direction. They can also accelerate electrons in the perpendicular direction through cyclotron resonance. Our work emphasizes the effectiveness of ECH waves as well as electron acoustic waves on regulating electron dynamics through heating of cold electrons and relaxation of beam electrons. Low energy electron beams, which are likely originated in the ionosphere \cite{Khazanov14:outflow, Artemyev20:jgr:feedback}, provide the free energy source for ECH wave excitation in Earth’s plasma sheet \cite{Zhang21:jgr:ECH}. Ionospheric electron outflows are accelerated and thermalized in parallel direction through Landau resonance with ECH waves, resulting in the trapping of ionospheric electron outflows in Earth’s plasma sheet. Such processes could potentially play an important role in the formation of field-aligned anisotropic electrons. Such electrons are frequently observed in Earth's magnetotail, in the energy range from a few eV to a few hundred eV \cite{Walsh11,Artemyev14:jgr,Walsh20}.     

ECH and electron acoustic waves are excited by field-aligned electron beam. Although electron acoustic waves are frequently observed to propagate in parallel directions with respect to the ambient magnetic field \cite{Gary&Tokar85,Gary87}, such waves are propagating at oblique wave normal angles in our simulation. The large temperature of the electron beam (100eV) and hence the small phase space density slope $\partial{f}/\partial{v_{\parallel}}$ suppresses the free energy for generation of parallel propagating electron acoustic waves through Landau resonance. Oblique propagating electron acoustic waves, however, can still be driven unstable through cyclotron resonance with the electron beam in our simulation. Under slightly different plasma parameters, the electron acoustic wave can be driven unstable through Landau resonance with the electron beam and propagate in the parallel direction \cite{Zhang21:pop:ECH}. In addition to the electron acoustic and ECH waves previously discussed, oblique whistler-mode waves can also be driven unstable by the electron beam \cite{Artemyev16:ssr, Li16}. These waves carry strong parallel electric fields and are quasi-electrostatic \cite{Agapitov13:jgr,Taubenschuss14,Artemyev15:natcom}. Under certain background plasma parameters, whistler-mode waves might couple to electron acoustic waves, making it difficult to distinguish between the two \cite{Vasko18:prl}. The relative importance between whistler-mode waves, electron acoustic waves and ECH waves under different plasma conditions remains an open question. In the future, it is important to investigate (1) the relative intensity of the whistler-mode waves, electron acoustic waves and ECH waves under different plasma conditions; (2) the different effects of whistler-mode waves, electron acoustic waves and ECH waves on modifying electron distribution functions and regulating electron dynamics; and (3) how the evolving electron distributions could affect wave saturation and wave properties. Direct comparison with wave observations can help better understand the electron sources and dynamics in a geophysical context, since plasma waves can be readily observed (with high time resolution and over large distances from their source) whereas electron distribution functions are often hard to measure fast, accurately, or in an unstable state.

Under electron distribution functions that represent typical plasma sheet environment, our simulation results have the potential to cast light on wave-particle interaction from observational point of view. Excited through cyclotron resonance with low energy electron beam, both moderately oblique ECH waves and electron acoustic waves are likely to be observed in plasma sheet. ECH waves have been frequently observed to be correlated with injections and dipolarization fronts in Earth's plasma sheet \cite{Liang11:ECH,Zhang14:ECH&DF,Zhang21:jgr:ECH}. Electrostatic solitary waves, which are often interpreted as the nonlinear structures formed during the excitation of electron acoustic waves, have also been observed near particle injections and dipolarization fronts \cite{Ergun09,Deng10,Malaspina18}. It is worth investigating the occurrence rates and intensities of beam-driven ECH wave and beam-driven electron acoustic wave and their relations to transient magnetotail phenomena such as injections and dipolarization fronts \cite{Runov09grl,Liu13:DF}. Both waves can scatter plasma sheet electrons into the loss cone and contribute to diffuse aurora precipitation \cite{Ni11:ECH,Ni11:plasma_sheet,Ni12:ECH,Zhang15:ECH,Vasko17:diffusion,Shen20:jgr:tds,Shen20:tds}. Furthermore, they carry strong parallel electric fields and can accelerate electrons of a sub-thermal energy range in the parallel direction through Landau resonance. Investigating the relation between beam-driven ECH waves and beam-driven electron acoustic waves can improve our understanding on the potential role of these two waves on affecting electron dynamics in the plasma sheet. 
\nocite{bookStix62,bookHelliwell65,Shklyar04}

\newpage
Table 1 is the initial electron distribution functions
\begin{table} [h!]
\centering
\begin{tabular}{ |p{4cm}|p{2.5cm}|p{2.5cm}|p{2.5cm}|p{2.5cm}| } 
  \hline
  & $n(cm^{-3})$ & $T_{\parallel}(eV)$ & $T_{\perp}/T_{\parallel}$ & $v_{drift}/v_{th}$ \\ 
  \hline
   Hot component & 0.5 & 1000 & 0.85 & 0 \\
  \hline
   Cold component & 0.05 & 1 & 1 & 0 \\
  \hline
   Beam component & 0.05 & 100 & 0.85 & 3 \\
  \hline
\end{tabular}
\label{table:1}
\end{table}
\clearpage

\acknowledgments
This work was supported from NASA grants 80NSSC22K1638, 80NSSC22K1634 and NASA contract NAS5-02099. Xin An acknowledges support by NASA grant 80NSSC22K1634. Anton Artemyev acknowledges support by NASA grant 80NSSC23K0413. Xiaojia Zhang acknowledges support by NASA Grant 80NSSC21K0729. We would like to acknowledge high-performance computing support from Cheyenne (doi:10.5065/D6RX99HX) provided by NCAR's Computational and Information Systems Laboratory, sponsored by the National Science Foundation.

\section*{Open Research} \noindent The simulation data have been archived on Zenodo: \cite{Zhang23JGR:zenodo} 

\bibliography{full}

\end{document}